\documentclass[aps,pra, twocolumn,showpacs,superscriptaddress]{revtex4-1}

\usepackage{graphicx}
\usepackage{SIunits}

\newcommand{\imag}{\mathrm{i}}
\newcommand{\R}{{\mathrm{R}}}
\renewcommand{\L}{{\mathrm{L}}}
\newcommand{\LHM}{\textsc{lhm}}
\newcommand{\RHM}{\textsc{rhm}}
\newcommand{\CS}{\textsc{cs}}
\newcommand{\HSS}{\textsc{hss}}

\begin{document}

\title{Impact of nonlocal interactions in dissipative systems: towards minimal-sized localized structures}

\author{Lendert \surname{Gelens}}
\affiliation{Department of Applied Physics and Photonics,
             Vrije Universiteit Brussel,
             Pleinlaan 2,
             B-1050 Brussel,
             Belgium}

\author{Guy \surname{Van der Sande}}
\affiliation{Department of Applied Physics and Photonics,
             Vrije Universiteit Brussel,
             Pleinlaan 2,
             B-1050 Brussel,
             Belgium}

\author{Philippe \surname{Tassin}}
\affiliation{Department of Applied Physics and Photonics,
             Vrije Universiteit Brussel,
             Pleinlaan 2,
             B-1050 Brussel,
             Belgium}
             
\author{Mustapha \surname{Tlidi}}
\affiliation{Optique non lin\'{e}aire th\'{e}orique,
             Universit\'{e} Libre de Bruxelles, CP\,231,
             Campus Plaine,
             B-1050 Bruxelles,
             Belgium}
             
\author{Pascal \surname{Kockaert}}
\affiliation{Optique et acoustique,
             Universit\'{e} Libre de Bruxelles, CP\,194/5,
             50~Av.~F.D.~Roosevelt, B-1050 Bruxelles, Belgium}

\author{Damia \surname{Gomila}}
\affiliation{Unidad de F\'isica Interdisciplinar (CSIC-UIB),
             Campus Universitat Illes Balears, 
             E-07122 Palma de Mallorca, Spain}
             
\author{Irina \surname{Veretennicoff}}
\affiliation{Department of Applied Physics and Photonics,
             Vrije Universiteit Brussel,
             Pleinlaan 2,
             B-1050 Brussel,
             Belgium}

\author{Jan Danckaert}
\affiliation{Department of Applied Physics and Photonics,
             Vrije Universiteit Brussel,
             Pleinlaan 2,
             B-1050 Brussel,
             Belgium}

\date{\today}

\begin{abstract}
In order to investigate the size limit on spatial localized structures in a nonlinear system, we explore the impact of linear nonlocality on their domains of existence and stability. Our system of choice is an optical microresonator containing an additional metamaterial layer in the cavity, allowing the nonlocal response of the material to become the dominating spatial process. In that case, our bifurcation analysis shows that this nonlocality imposes a new limit on the width of localized structures going beyond the traditional diffraction limit.

\end{abstract}

\maketitle

Localized structures, belonging to the class of dissipative structures found in systems far from equilibrium, are stationary peaks in one or more spatiotemporal dimensions. They occur in diverse fields of nonlinear science, such as chemistry \cite{Pearson-1993,Lee-1993}, plant ecology \cite{Lejeune-2006}, gas discharge systems \cite{Muller-1999}, fluids \cite{Thual-1990} and optics \cite{Scroggie-1994, Taranenko-1997,2002Natur.419..699B,Bortolozzo-2006,Trillo-2001,Rosanov-2002,Mandel-2004}. The presence of localized structures has been demonstrated in systems exhibiting a nonlocal response, such as in models for population dynamics \cite{2004PhRvE_Hernandez}, neural networks \cite{coombes05} and nonlocal optical materials \cite{Krolikowski-2004, Skopin-2007}.  Although the formation and stability of these localized structures must be influenced by these nonlocal effects, the main effects of nonlocality are typically obscured by the presence of a stronger diffusion and/or diffraction process. Thanks to recent advances in the fabrication of metamaterials, it is now possible to conceive a nonlinear system under conditions such that the dynamical properties of localized structures are dominated by the nonlocal response. In this work, it is our aim to analyze the influence of this nonlocality on the nonlinear dynamical properties of two-dimensional (2D) spatial structures. 

Localized structures are relatively well understood in one transverse dimension \cite{Coullet-2000}, whereas an analytical analysis in two transverse dimensions is still largely unexplored and most of the results are obtained by numerical simulations. In this work, we will focus on the particular system of controllable 2D localized structures in optical cavities, also referred to as cavity solitons (\CS{}s).
Their formation can be attributed to the balance between nonlinearities due to light-matter interaction, transport processes (diffusion and/or diffraction), and dissipation \cite{Taranenko-1997,2002Natur.419..699B,1997PhRvL..79.2042B,Tlidi-1994}. These bright spots have been proposed for information encoding and processing \cite{2002Natur.419..699B,1997PhRvL..79.2042B,Hachair-2004}. Decreasing the size of \CS{}s would be advantageous for these applications, while also being of fundamental interest. In optical devices, the spatial dimension of \CS{}s is generally restricted by diffraction, which imposes the diameter to be of the order of $\sqrt{\mathcal{D}}$, where $\mathcal{D}=l\mathcal{F}/(\pi{k})$ is the diffraction coefficient, with $k$ the wavenumber of the beam, and $l$ and $\mathcal{F}$ the length and the finesse of the resonator. For example, in Ref.~\cite{2002Natur.419..699B}, \CS{}s at $0.5{\micro\meter}$ have a transverse size that reaches the diffraction limit of $\unit{10}{\micro\meter}$.

The use of transverse index modulation has been proposed to overcome the diffraction limit of mid-band \CS{}s in a certain class of resonators \cite{2003PhRvL..91e3901S}.
These localized structures have a large intrinsic transverse velocity that prevents its use for many practical applications.
Recently, a different strategy was motivated by the progression of the fabrication of left-handed materials (\LHM) towards optical frequencies  \cite{smith2004, Shalaev-2007}, and the proposition of nonlinear \LHM{} \cite{Zharov-2003, Lapine-2003}. In
Refs.~\cite{2006PhRvA..74c3822K,Tassin-2006}, Kockaert~\textit{et al.}\ study the possibility of altering the strength of diffraction by inserting a layer of \LHM{}, in addition to a layer of right-handed material (\RHM), in the cavity of an optical microresonator, and they show how to reduce the diffraction coefficient to arbitrarily small values. As the soliton width scales with the square root of
the diffraction coefficient, this method potentially allows for sub-diffraction-limited \CS{}s. Unlike with the use of photonic crystals, this technique works in principle for all types of \CS{}s and microresonators. Although the diffraction limit can be encompassed in this way, one can reasonably expect that the sub-wavelength structure of \LHM{}s will impose a new size limit on \CS{}s. 
Here, we will show that the inherent nonlocality of \LHM{}s will significantly change the properties of \CS{}s when diffraction is tuned down, and a new size limit of \CS{}s --- now imposed by nonlocality --- will be revealed.

We consider a microresonator driven by a coherent optical beam. In each roundtrip, the light passes through two adjacent nonlinear
Kerr media: a \RHM{} and a \LHM. It has been shown in Ref.~\cite{2006PhRvA..74c3822K} that
the evolution of the electric field in this microresonator is governed by the well-known Lugiato-Lefever (LL) equation \cite{LL-1987}, with the diffraction coefficient  $\mathcal{D}$ given by
\begin{equation}
\mathcal{D} =
\frac{\lambda\,\mathcal{F}}{2\pi^2}\left(\frac{l_\R}{n_\R}-\frac{l_\L}{\vert{n_\L}\vert}\right). \label{Eq:DiffrCoeff}
\end{equation}
$n_\R$, $n_\L$, $l_\R$ and $l_\L$ are the indices of refraction and the
lengths of the \RHM{} and the \LHM, respectively. By changing $l_\R$ and $l_\L$, $\mathcal{D}$ can be engineered to ever smaller, positive diffraction coefficients. From the LL equation, one can estimate the diameter of localized solutions to be
\begin{equation}
\Lambda_0 = 2 \pi \sqrt{\frac{\mathcal{D}}{2-\Delta}} \label{Eq:Lambda0} ,
\end{equation}
with $\Delta$ the cavity detuning.
Therefore, \CS{}s will become infinitely small when $\mathcal{D}$ tends to zero. But in that case, higher order effects will start to dominate the spatial dynamics. Indeed, from the derivation of the LL equation, one can show that the inherent nonlocality of the nanostructured metamaterial comes into play when diffraction becomes negligible. The nonlocality in these materials is still largely unexplored. When addressing the cavity with an optical beam, the induced fields in the sub-wavelength resonators will also couple to neighboring resonators, providing a nonlocal response. We assume this nonlocal response to be weak, because the coupling to the nearest neighbor resonators is important.

Under the same approximations under which the LL equation is valid, i.e., slowly varying envelope approximation, weak nonlinearity and a nearly resonant cavity, we find that the nonlocality comes in as follows:
\begin{eqnarray}
\frac{\partial A}{\partial{t}}
=
-(1+\imag\Delta )A+A_\mathrm{in}+\imag\,\vert A \vert^2\,A
+\imag\mathcal{D}\nabla_\perp^2 A \nonumber\\
+ \imag \int\!\!\!{\int{\theta(\textbf{r}_{\perp}-\textbf{u})A(\textbf{u})\rm{d}\textbf{u}}}.
\label{Eq:LL2}
\end{eqnarray}
The kernel function $\theta$, describing the nonlocal response of the linear \LHM, effectively couples the electric field at different positions. The details of the derivation of Eq.~(\ref{Eq:LL2}) will be published elsewhere. When the nonlocality is weak, the last term of Eq.~(\ref{Eq:LL2}) can be expanded in a series of spatial derivatives of $A$, and, taking into account the rotational invariance of the system, we find
\begin{eqnarray}
\int\!\!\!{\int{\theta(\textbf{r}_{\perp}-\textbf{u})A(\textbf{u})\rm{d}\textbf{u}}}
\simeq \theta_0 A + \theta_1 \nabla_\perp^2 A +\theta_2 \nabla_\perp^4 A.
\label{Eq::approximation}
\end{eqnarray}
The first two terms in Eq.~(\ref{Eq::approximation}) only change the value of the detuning and diffraction coefficients, respectively. In what follows, we absorb their contribution into the parameters $\Delta$ and $\mathcal{D}^{(1)}$, keeping the same notation. We finally arrive at
\begin{equation}
\frac{\partial A}{\partial{t}}  = -(1+\imag\Delta )A+A_\mathrm{in}+\imag\,\vert{A}\vert^2\,A +\imag\mathcal{D}^{(1)}\nabla_\perp^2 A + \imag\mathcal{D}^{(2)}\nabla_\perp^4 A.
\label{Eq:LL3}
\end{equation}

Eq.~(\ref{Eq:LL3}) is similar to the LL equation, and has the same homogeneous steady state (\HSS) solutions $A_{\mathrm{s}}$. In this work, we want to study the localized structures arising from the modulational instability of the homogeneous solution.
Therefore, we restrict ourselves to the monostable regime ($\Delta < \sqrt{3}$).
We have performed a stability analysis by linearizing Eq.~(\ref{Eq:LL3}) around
the steady state solution, and seeking for the deviation in the form
\(A=A_{\mathrm{s}}+\delta{}A\,\mathrm{\ \exp (}\imag \,\mathbf{k}\cdot
\mathbf{r}+\lambda {t),}\) with \(\mathbf{k}=(k_{x},k_{y})\) and
\(\mathbf{r}=(x,y)\). The marginal stability is given by
\begin{equation}
\mathcal{D}^{(1)} k_\mathrm{m}^2 - \mathcal{D}^{(2)} k_\mathrm{m}^4 = 2 \vert A_\mathrm{s} \vert^2 - \Delta \pm \sqrt{\vert A_\mathrm{s} \vert^4 - 1}. \label{Eq:MarginalStability}
\end{equation}
The modulational instability (MI) depends strongly on the parameter $\eta$, defined as the nonlocality to diffraction ratio $\eta = \mathcal{D}^{(2)} / {\mathcal{D}^{(1)}}^2$.
\begin{figure}
\includegraphics[clip]{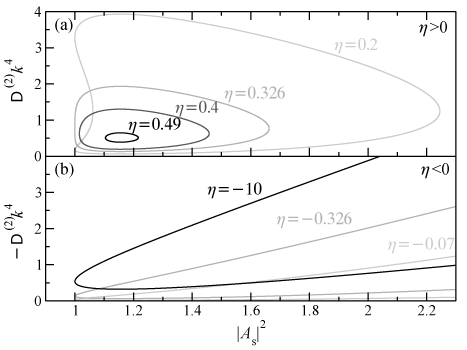}
\caption{Influence of the nonlocality to diffraction ratio $\eta$ on the marginal stability curves as given by Eq.\ (\ref{Eq:MarginalStability}). The wavevectors inside the curves destabilize the homogeneous steady state solution of Eq.\ (\ref{Eq:LL3}). $\Delta = 1.23$. (a) Positive $\eta$. (b) Negative $\eta$.}
\label{Fig::MarginalStability}
\end{figure}
In Fig.~\ref{Fig::MarginalStability}, the marginal stability curves are shown
for several values of $\eta$.

For $\eta > 0$ [see Fig.~\ref{Fig::MarginalStability}(a)], these curves have the
form of a cardioid, which contains the unstable wavevectors. A first MI arises
with two critical wavenumbers at an intensity of $\vert A_\mathrm{s} \vert^2 =
1$. This leads to the formation of complex patterns with two
wavevector components. Stability of the \HSS{} is recovered at higher background
intensities. As $\eta$ increases, the cardioid form evolves into an elliptical
shape with equal critical wavenumbers at both sides:
$\mathcal{D}^{(2)}k_\mathrm{m}^4 = 1 / 4\eta$. From this wavenumber, we can
estimate the typical width of localized structures to be of the order of
\begin{equation}
\Lambda_+ = 2 \pi \sqrt[4]{ 4 \mathcal{D}^{(2)} \eta} . \label{Eq:LambdaPlus}
\end{equation}
In the limit of small diffraction, Eq.~(\ref{Eq:LambdaPlus}) predicts that the
width of \CS{}s increases when  $\mathcal{D}^{(1)}$ decreases, while
Eq.~(\ref{Eq:Lambda0}) indicates a decreasing width. Therefore, there must exist a value of the diffraction
strength $\mathcal{D}^{(1)}$ for which the \CS\ is of minimal width. We will use a numerical method to determine
this optimum. However, we want to point out that this minimal size \CS{} can be
unstable. Our calculations will also provide insight in this matter. Also note that, when $\eta > 1 / 4 \left( {\sqrt{3}-\Delta} \right)$,
the unstable region contained in the marginal stability curve disappears,
effectively stabilizing the \HSS.

For $\eta < 0$ [see Fig.~\ref{Fig::MarginalStability}(b)], the modulational instability occurs again for intensities $\vert A_\mathrm{s} \vert^{2} > 1$, but here the \HSS{} is not stabilized for high input fields. At the critical point, we can derive the following estimate for the \CS{} width:
\begin{equation}
\Lambda_- = \frac{2 \pi \sqrt[4]{4 \mathcal{D}^{(2)}\eta}} {\sqrt{ 1+ \sqrt{1-4 (2-\Delta)\eta} }}. \label{Eq:LambdaMin}
\end{equation}
Consequently, when $\eta < 0$, the \CS{} width will still decrease with the
diffraction strength $\mathcal{D}^{(1)}$, but due to the nonlocality the width will
saturate at $\Lambda_\mathrm{lim} = 2 \pi \sqrt[4]{-\mathcal{D}^{(2)}/(2-\Delta)}$. Again, the possibility that the \CS{}s
become unstable before reaching this limit exists, emphasizing the need
to check the stability of these structures.

Our numerical analysis relies on a calculation of the stationary localized solutions of Eq.~(\ref{Eq:LL3}) using a Newton-Raphson method \cite{2002JOSAB..19..747F,2005PhRvL..94f3905G, Damia_PRE}. The radial form of this equation is discretized, from which a set of coupled nonlinear equations is obtained. Since the equation is linear in the spatial derivatives, the diffraction term can be computed in the spatial Fourier space. Zero derivatives at the boundaries are imposed. This approach is extremely accurate and generates the Jacobian.

\begin{figure}
\includegraphics[clip]{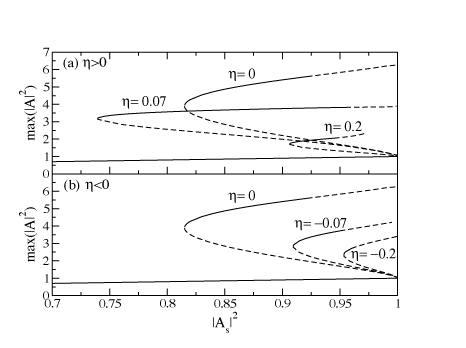}
\caption{Maximal intensity of the cavity soliton vs.\ the background intensity for different values of $\eta$. Solid lines indicate stable structures, whereas dashed lines correspond to unstable solutions. $\Delta = 1.23$. (a) Positive $\eta$. (b) Negative $\eta$.}
\label{Fig::Bifurcation}
\end{figure}

In Fig.~\ref{Fig::Bifurcation}, \CS{} bifurcation diagrams are shown for different values of $\eta$. In the absence of nonlocality ($\eta = 0$), as previously studied in Refs.\ \cite{2002JOSAB..19..747F,2005PhRvL..94f3905G}, a \CS{} branch emerges subcritically from the \HSS{} at $\vert A_\mathrm{s} \vert^2 = 1$. The negative slope part of this branch corresponds to unstable \CS{}s; the positive slope higher branch is stable for low values of $\vert A_\mathrm{s} \vert^2$.
At higher background intensities the intensity peaks become higher and narrower, attributed to the more prominent self-focusing effect, and at a certain point they become again unstable with respect to azimuthal perturbations.   
For  $\eta > 0$ [Fig.~\ref{Fig::Bifurcation}(a)], we observe that the branches extend to smaller  $\vert A_\mathrm{s} \vert^2$ with increasing $\eta$, while the domain of stability is enlarged. This trend is reversed from a certain $\eta$ on, and the stability range decreases. Also note that due to nonlocality, which physically tends to spread out the inensity to neighboring points, the peak intensity of \CS{}s drops. For $\eta < 0$ [Fig.~\ref{Fig::Bifurcation}(b)], the branches move monotonically to higher background intensity and to lower peak intensity with stronger nonlocality. Again note that the stability range is reduced considerably.

\begin{figure}
\includegraphics[clip]{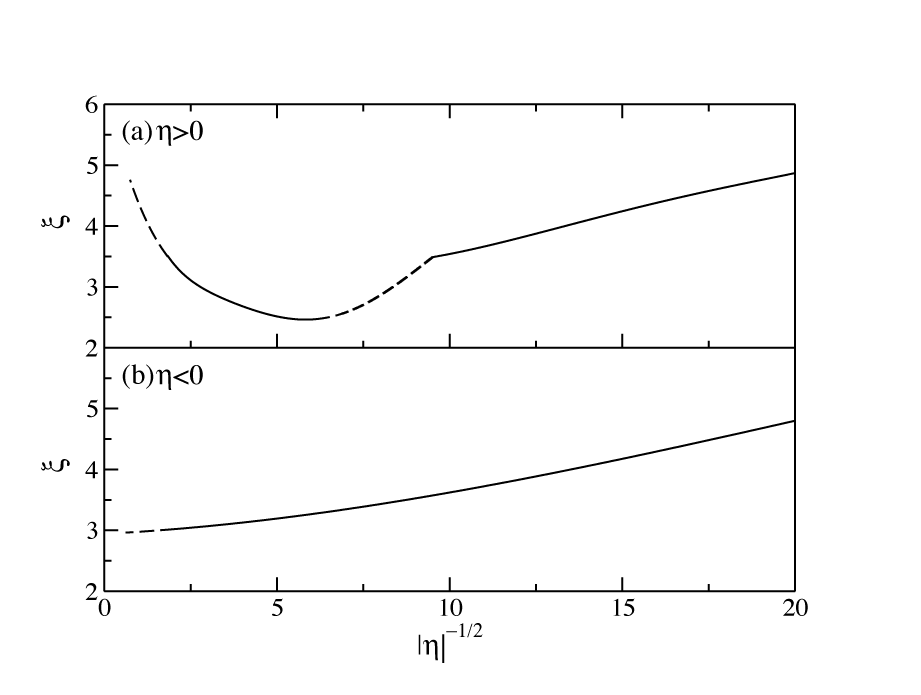}
\caption{Renormalized full width at half maximum $\xi$ vs.\ diffraction strength ($|\eta|^{-1/2}$). The width $\xi$ is shown of the higher branch \CS{} at the change of stability or of the \CS{} at the saddle-node bifurcation when no stable \CS{}s exist  Stable \CS{}s are indicated by solid lines, whereas the unstable \CS{}s are in dotted lines. $\Delta = 1.23$. (a) Positive $\eta$. (b) Negative $\eta$.}
\label{Fig::CSWidth}
\end{figure}

To find the minimal \CS{} size as discussed above, we have investigated the scaling of the renormalized full width at half maximum ($\xi = d_\mathrm{FWHM} / \sqrt[4]{\vert \mathcal{D}^{(2)} \vert}$) of the \CS{}s with $\eta$ in Fig.~\ref{Fig::CSWidth}. Note that this renormalization enables us to obtain a general result only depending on the detuning $\Delta$. For each value of $\eta$, we indicate the width $\xi$ of the upper branch \CS{}s at the change of stability, when stable \CS{}s exist, while plotting the $\xi$ at the saddle-node bifurcation when the entire upper branch is unstable. For $\eta > 0$ [Fig.~\ref{Fig::CSWidth}(a)], one can distinguish two regions of stable \CS{}s, one in which the width decreases with decreasing diffraction strength, and the other with the inverse effect. One can identify the minimal width $\xi_\mathrm{min}$ and its corresponding optimal $\eta_\mathrm{opt}$ in the leftmost region. For $\eta < 0$ [Fig.~\ref{Fig::CSWidth}(b)], the width decreases monotonically with decreasing diffraction strength, until stability of the \CS{}s is lost. At this point, the minimal width $\xi_{min}$ is obtained. Our numerical calculations thus qualitatively confirm the prediction by the linear stability analysis given above.

\begin{figure}
\includegraphics[clip]{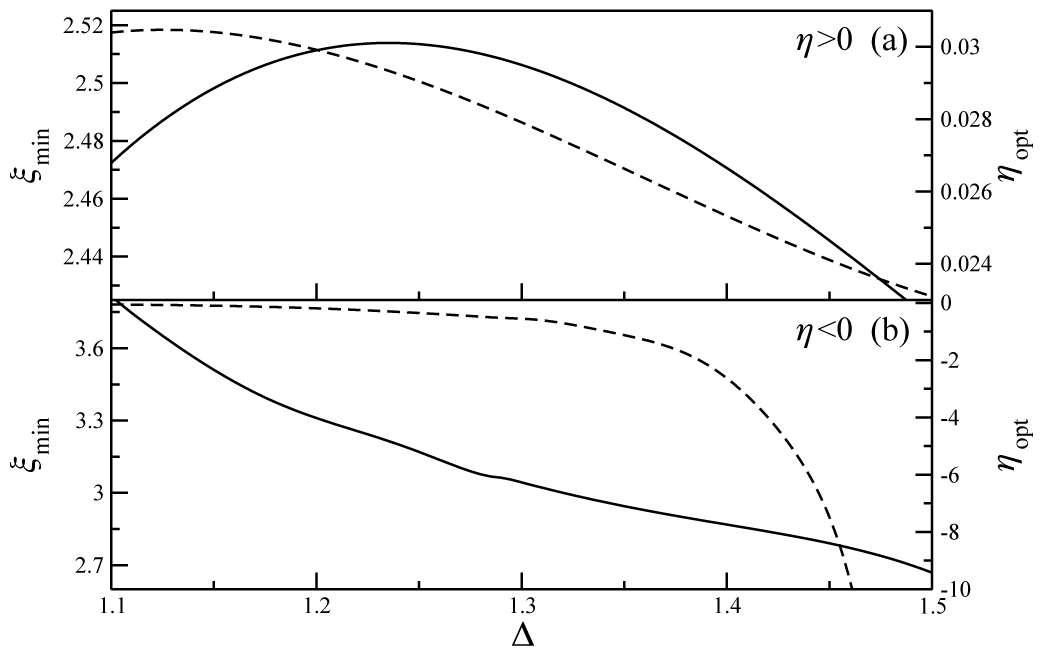}
\caption{Minimal \CS{} width $\xi_{min}$ (solid lines, left axis) and optimal $\eta_{opt}$ (dashed lines, right axis) vs.\ detuning $\Delta$.(a) Positive $\eta$. (b) Negative $\eta$.}
\label{Fig::OptimalWidth}
\end{figure}

We have repeated this procedure for several values of the detuning. This gives us the optimal parameter set that produces minimal size \CS{}s given a certain nonlocality $\mathcal{D}^{(2)}$. Fig.~\ref{Fig::OptimalWidth} summarizes the main results. For $\eta > 0$ [Fig.~\ref{Fig::OptimalWidth}(a)], one can see that the minimal width $\xi_\mathrm{min}$ varies only by a few percent in this range of the detuning. The smallest $\xi_\mathrm{min}$ can be obtained for higher values of the detuning $\Delta$, and this with smaller values of $\eta_{opt}$. For the corresponding optimal values of $\eta$, $\mathcal{D}^{(2)}$ remains smaller than ${\mathcal{D}^{(1)}}^2$, so the formation of the minimal width \CS{}s is still dominated by diffraction effects. For $\eta < 0$ [Fig.~\ref{Fig::OptimalWidth}(b)], the minimal width again decreases with increasing values of the detuning, but changes more strongly than for positive $\eta$. Note that this time absolute values of $\eta_{opt}$ need to be larger to obtain the smallest \CS{}s. Hence, for $\eta < 0$, it is the nonlocality that has the largest influence on the formation of minimal size \CS{}s.

In conclusion, we have studied the impact of nonlocality on the nonlinear dynamical properties of two-dimensional spatially localized structures. Our study confirms the possibility to reduce the size of cavity solitons beyond the diffraction limit by using left-handed materials. However, we have shown that the nonlocal interaction between the optical field and the material not only hinders this size reduction, but also alters the stability of cavity solitons in a manifest way, imposing a new limit on their size.

\begin{acknowledgments}
This work was supported by the Belgian Science Policy Office under grant No.\ IAP-VI10. LG and PT are PhD Fellows and GV is a Postdoctoral Fellow of the Research Foundation - Flanders (\textsc{FWO} - Vlaanderen). MT is a Research Associate of the Fonds National de
la Recherche Scientifique (\textsc{FNRS}). DG acknowledges financial support from MEC (Spain) and FEDER: Grant TEC2006-10009 (PhoDeCC). 
\end{acknowledgments}


\end{document}